\documentstyle[amsfonts,12pt]{article}  
\advance\hoffset by -7mm  
\makeatletter
\@addtoreset{equation}{section}
\makeatother

\renewcommand{\title}[1]{\null\vspace{25mm}

\noindent{\Large{\bf #1}}\vspace{10mm}

\noindent {\large By }}
\newcommand{\authors}[1]{\noindent{\large #1}\vspace{3mm}

}
\newcommand{\address}[1]{\noindent #1\vspace{5mm}

}
\renewcommand{\abstract}[1]{\vspace{19mm}

\noindent{\small{\em Abstract.} #1}\vspace{2mm}

} 
\def\greaterthansquiggle{\raise.3ex\hbox{$>$\kern-.75em\lower1ex\hbox{$\sim$}}}
\def\lessthansquiggle{\raise.3ex\hbox{$<$\kern-.75em\lower1ex\hbox{$\sim$}}}
\newcommand{\beq}{\begin{equation}}

\newcommand{\eeq}{\end{equation}}
\newcommand{\beqa}{\begin{eqnarray}}
\newcommand{\eeqa}{\end{eqnarray}}
\newcommand{\beqan}{\begin{eqnarray*}}
\newcommand{\eeqan}{\end{eqnarray*}}
\newcommand{\ba}{\begin{array}}
\newcommand{\ea}{\end{array}}

\newcommand{\communication}{{communications}}
\newcommand{\communications}{{communications}}

\def\nz{\ifmmode {I\hskip -3pt N} \else {\hbox {$I\hskip -3pt N$}}\fi}
\def\zz{\ifmmode {Z\hskip -4.8pt Z} \else
       {\hbox {$Z\hskip -4.8pt Z$}}\fi}
\def\qz{\ifmmode {Q\hskip -5.0pt\vrule height6.0pt depth 0pt
       \hskip 6pt} \else {\hbox
       {$Q\hskip -5.0pt\vrule height6.0pt depth 0pt\hskip 6pt$}}\fi}
\def\rz{\ifmmode {I\hskip -3pt R} \else {\hbox {$I\hskip -3pt R$}}\fi}
\def\cz{\ifmmode {C\hskip -4.8pt\vrule height5.8pt\hskip 6.3pt} \else
       {\hbox {$C\hskip -4.8pt\vrule height5.8pt\hskip 6.3pt$}}\fi}
\def\au{{\setbox0=\hbox{\lower1.36775ex\hbox{''}\kern-.05em}\dp0=.36775ex\hskip0pt\box0}}
\def\ao{{}\kern-.10em\hbox{``}}

\newtheorem{Theorem} {Theorem} [section]

\newtheorem{Lemma} [Theorem] {Lemma}
\newtheorem{Proposition} [Theorem] {Proposition}

\newcommand{\R}{{\Bbb R}}

{\catcode `\@=11 \global\let\AddToReset=\@addtoreset}
\AddToReset{equation}{section}

\newcommand{\subjclass}[1]{}

\newcommand{\proof}{{\sc Proof:}\ }

\def\scri{\hbox{${\cal J}$\kern -.645em {\raise
      .57ex\hbox{$\scriptscriptstyle (\ $}}}}
\newcommand{\Scri}{\scri}

\newcommand{\cO}{{\cal O}}

\newcommand{\eq}[1]{(\ref{#1})}

\newcommand{\commentout}[1]{}

\newcommand{\ee}{\end{equation}}
\newcommand{\bea}{\begin{eqnarray}}
\newcommand{\eea}{\end{eqnarray}}
\newcommand{\beaa}{\begin{eqnarray*}}
\newcommand{\eeaa}{\end{eqnarray*}}

\newcommand{\Sext}{\Sigma_{\mathrm ext}}
\newcommand{\Mext}{M_{\mathrm ext}}
\newcommand{\doc}{\langle\langle M_{\mathrm ext}\rangle\rangle}

\begin{document}
\title{Uniqueness of stationary, electro--vacuum black holes \\[2mm]
revisited}
\authors{Piotr T. Chru\'sciel}
\address{D\'epartement de Math\'ematiques, Facult\'e des Sciences\\
  Parc de Grandmont, F37200 Tours, France}
%\maketitle

\abstract{%
  In recent years there has been some progress in the understanding of
  the global structure of stationary black hole space--times. In this
  paper we review some new results concerning the structure of
  stationary black hole space--times. In particular we prove a
  corrected version of the ``black hole rigidity theorem'', and we
  prove a uniqueness theorem for static black holes with degenerate
  connected horizons. 
  This paper is an expanded version of a lecture
  given at the Journ\'ees relativistes in Ascona, May 1996.
}
%\end{abstract}

\section{Introduction}

\label{sec:intro}

The theory of black hole space--times seems to be one of the most
elegant chapters of classical general relativity, that makes use of a
whole range of techniques: global causality theory, the theory of
partial differential equations, various aspects of Riemannian and
Lorentzian geometry, etc. The foundations of the theory have been laid
by Israel \cite{Israel:uniqueness}, Carter \cite{CarterlesHouches},
Hawking \cite{Ha1,HE}, and our knowledge of those space--times has
been considerably expanded through the work in \cite
{Mazur,Ruback,bunting:masood,Sudarsky:wald} and others. A recent
treatise which covers in detail various aspects of the theory is
\cite{Heusler:book}.  In spite of the considerable amount of work
involved, several questions still remain to be investigated ({\em
  cf.\/}~\cite{Chnohair} for a recent review). In the last few years
some of the open problems raised in that last reference have been
solved, and in this review we shall shortly discuss some of those
developments. I will also take this opportunity to give a complete
proof of a corrected version of the so--called ``rigidity theorem'' for
stationary black holes, as well as some steps of the proof of a
uniqueness theorem for degenerate Reissner--Nordstr\"om black holes. We
shall mostly concentrate on electro--vacuum space--times; a discussion
of various results on some other models can be found in \cite
{Heusler:book,Bizon:bhreview,Garyonbh}. The reader is also referred to
\cite {SudarskyNunez} for an interesting new result concerning
spherically symmetric ``hairy'' black holes.

Let us start by recalling that a space--time\footnote{ Throughout this
paper the term space--time denotes a smooth, paracompact, connected,
orientable and time--orientable Lorentzian manifold.} is called {\em
stationary } if there exists a Killing vector field $X$ which
approaches $\partial /\partial t$ in the asymptotically flat region
{\em and } generates a one parameter groups of isometries (for the
purposes of this paper space--times will be assumed to be
asymptotically flat.) A space--time is called {\em static } if it is
stationary and if there exists an isometry which changes time
orientation. A space--time is called {\em axisymmetric } if there
exists a Killing vector field $Y$ which behaves like a {\em rotation}
in the asymptotically flat region {\em and} generates a one parameter
group of isometries, in the following (standard) sense: in an
asymptotically Minkowskian coordinate system the partial derivatives
of the Killing vector field asymptote to those of a rotational Killing
vector field of Minkowski space--time, and all orbits of $ Y$ are,
say, $2\pi $ periodic. It can be shown that this implies that there
exists an axis of symmetry, that is, a set on which the Killing vector
vanishes ({\em cf.\ e.g.\/} \cite[Prop. 2.4]{ChBeig2}). That last
property is often imposed as a part of the definition of
{axisymmetry}.

It is worthwhile emphasizing that the above notions include the property
that the orbits of the relevant Killing vectors are {\em  complete}. This
is a non--trivial requirement which  plays an important role in the
theory, and several steps of the proofs are wrong without that hypothesis.
Hence, the question of completeness of the orbits, which we will discuss
shortly in Section \ref{sec:completeness} below, is not only a question of
mathematical purity, but is critical in several considerations.

It is widely expected that the Kerr--Newman black holes, together perhaps
with the Majumdar--Papapetrou ones, exhaust the set of appropriately regular
stationary electro--vacuum black holes. (An important question is, what does
one mean by ``appropriately regular''.) This expectation is based on the
following facts, which have been proved under various restrictive
hypotheses, some of which we shall present in detail below:

\begin{enumerate}
\item  The ``rigidity theorem'': non--degenerate\footnote{The condition
of non--degeneracy is only needed here to be able to infer staticity,
when the Killing vector field that asymptotes to $\partial/\partial t$
is tangent to the event horizon.}  stationary analytic electro--vacuum
black holes are either {\em static}, or {\em axisymmetric
(cf. }Theorem \ref {T:rigidity2} below).

\item  {} The Reissner--Nordstr\"{o}m non--degenerate black holes exhaust the
family of {\em static} non--degenerate electro--vacuum black holes.

\item  {} The non--degenerate Kerr--Newman black holes exhaust the
family of non--degenerate {\em   stationary--axisymmetric\/} electro--vacuum
 black holes ({\em  cf.\/} Theorem \ref{T:Weinstein} below).
\end{enumerate}

Clearly, the rigidity theorem plays a key role here, reducing the
classification problem of stationary black holes to that of classification
of the static and of the axisymmetric ones. (More precisely, that theorem
would play a key role if the analyticity condition there were removed.) It
turns out that this theorem is wrong as stated in \cite{HE}: in \cite
{Ch:rigidity} a space--time has been constructed which satisfies all the
hypotheses spelled--out in \cite{HE} and which is {\em not\/} axisymmetric.
The construction consists in gluing together, in an appropriate way, two
copies of the Kerr space--time, so that the isometry group of the resulting
space--time is only ${\Bbb R}$: while there are still two linearly
independent Killing vector fields globally defined, only one of them has
complete orbits. This can be considered as a rather mild counter--example,
but its existence shows that there is a potentially dangerous error in the
proof of the theorem. We shall show in Section \ref{sec:rigidity} below that
some form of rigidity, weaker than that claimed in \cite{HE}, can be
obtained. Let us start by presenting some recent results which will be used
in the proof.

\section{The topology of black hole space--times}

\label{sec:topology}
One of the steps of the proof of the rigidity theorem (both in
\cite{HE} and below) consists in proving that connected components of
black--hole event horizons must have ${\Bbb R}\times S^2$ topology.
Moreover one also needs a simply--connected domain of outer
\communication. Both claims are insufficiently justified in \cite{HE}.
The first complete proof of that result has been given in
\cite{ChWald}. The results of that reference have been generalized by
Galloway, as follows: Consider a space--time which is asymptotically
flat at null infinity in the sense of the definition given on p.~282 of
\cite{Waldbook}. We shall say that $\Scri$ satisfies {\em the
  regularity condition\/} if 
\begin{enumerate}
\item There exists a neighborhood $\widehat{\cal O}$ of $\Scri^+ \cup
  \Scri^-$ which is simply connected.
\item{} There exists a neighborhood ${\cal O}\subset\widehat{\cal O}$
  of $\Scri^-$ such that for every compact set $K\subset {\cal
O}\setminus \Scri^-$, the
  set $\partial J^+(K;\bar M)\cap\Scri^+$ is not empty.
\end{enumerate} 
We have the following result of Galloway, that does
{\em not\/} assume stationarity\cite{galloway-topology}:

\begin{Theorem}[G. Galloway, 96]
\label{T:galloway} Consider a space--time which is asymptotically flat
at null infinity and which has a $\Scri$ which satisfies the
regularity\footnote{The regularity condition of $\Scri$ can be
  replaced by other conditions, {\em cf.\ 
    e.g.\/}~\cite{Galloway:fitopology}.  While the regularity
  condition given here is not spelled out in detail in
  \cite{FriedmanSchleichWitt} and in \cite{Galloway:fitopology}, some
  form of causal regularity of $\Scri$ is needed for the arguments
  given there to go through. I am grateful to G.~Galloway for
  discussions concerning this point.} condition. Suppose moreover that
the Ricci tensor satisfies the null convergence condition:
\begin{equation}
R_{\mu \nu }X^\mu X^\nu \ge 0\quad \mbox{  for all null vectors $X^\mu$.}
\label{energy}
\end{equation}
Then every globally hyperbolic domain of outer \communication\ is simply
connected.
\end{Theorem}

The hypothesis of global hyperbolicity of the d.o.c. above is rather
reasonable; it should be pointed out that it necessarily holds if the
space--time itself is globally hyperbolic. All the above mentioned results
are based on the {\em topological censorship theorem\/} of Friedman,
Schleich and Witt \cite{FriedmanSchleichWitt}. Further related results can
be found in \cite
{Jacobson:venkatarami,Woolgar:BCP,galloway:woolgar,Galloway:fitopology}.

To be able to apply this result to stationary black hole space--times
we need to verify that the regularity condition of $\Scri$ is satisfied.
Recall, first, that there are two ways of defining the domain of outer
\communication\  for a stationary space--time, the one given in
\cite{ChWald1}, and the standard
one using $
\hbox{${\cal J}$\kern -.645em {\raise
      .57ex\hbox{$\scriptscriptstyle (\ $}}}$ ({\em cf.\ e.g.\/} \cite
{HE,Waldbook}). (Those definitions are equivalent for, say, electro--vacuum
stationary space--times, but the former seems to be more convenient for many
purposes.) The approach of \cite{ChWald1} proceeds as follows: A
space--like hypersurface $\Sigma _{\mathrm{ext}}$ diffeomorphic to
      ${\Bbb R}^3$ minus a 
ball will be said asymptotically flat if the fields $(g_{ij},K_{ij})$
induced on $\Sigma _{\mathrm{ext}}$ by the space--time metric satisfy
the fall--off 
conditions 
\begin{equation}
|g_{ij}-\delta _{ij}|+r|\partial _\ell g_{ij}|+\cdots +r^k|\partial _{\ell
_1\cdots \ell _k}g_{ij}|+r|K_{ij}|+\cdots +r^k|\partial _{\ell _1\cdots \ell
_{k-1}}K_{ij}|\le Cr^{-\alpha }\ ,  \label{falloff}
\end{equation}
for some constants $C$, $\alpha >0$. An asymptotically flat hypersurface $
\Sigma $ (with or without boundary) will be said to have compact interior if 
$\Sigma $ is of the form $\Sigma _{int}\cup \Sigma _{\mathrm{ext}}$,
with $\Sigma _{int}$ compact. Let $X$ be a Killing vector field which
asymptotically 
approaches the unit normal to $\Sigma _{\mathrm{ext}}$. Passing to a
subset of $\Sigma _{\mathrm{ext}}$ we can without loss of generality
assume that $X$ is timelike on $ \Sigma _{\mathrm{ext}}$. If $\phi _t$
denotes the one parameter group of isometries 
generated by $X$, then an exterior four--dimensional asymptotically flat
region can be obtained by moving $\Sigma _{\mathrm{ext}}$ around with
the isometries;  
\begin{equation}
M_{\mathrm{ext}}=\cup _{t\in {\Bbb R}}\phi _t(\Sigma _{\mathrm{ext}})\
.  \label{mext} 
\end{equation}
The {\em domain of outer \communications\/} (d.o.c.) associated with $\Sigma
_{\mathrm{ext}}$ or with $M_{\mathrm{ext}}$ is then defined as 
\begin{equation}
\langle \langle M_{\mathrm{ext}}\rangle \rangle
=J^{+}(M_{\mathrm{ext}})\cap J^{-}(M_{\mathrm{ext}})\ . 
\label{doc}
\end{equation}
The black--hole event horizon ${\cal B}$ associated with the asymptotic end $
\Sigma _{{\em ext}}$ or with $M_{{\em ext}}$ is defined as 
\begin{equation}
{\cal B}=M\setminus J^{-}(M_{\mathrm{ext}})\ .  \label{bh}
\end{equation}
For electro--vacuum space--times one can now attach to
$M_{\mathrm{ext}}$ a null conformal boundary using the prescription
given in \cite[Appendix]{Damour:schmidt} --- the conditions needed for
that construction are satisfied by \cite{Simon:elvac}. It is
straightforward to check that the regularity condition is satisfied by
this completion (with ${\cal O}=M_{\mathrm{ext}}\cup \Scri^-$,
$\widehat{\cal O}={\cal O}\cup \Scri^+$).

The spherical topology theorem is essentially a Corollary of Theorem
\ref {T:galloway}. More precisely, let us consider a space--like
hypersurface $\Sigma$ which has a boundary on the event horizon. We
wish to show that this boundary is a finite collection of spheres.
Clearly some further hypotheses are needed for this result to hold.
Let us start with some terminology.  Consider a point $p$ in the
asymptotically flat region $M_{{\mathrm{ext}}}$ so that, in
particular, $X^a$ is timelike at $p$.  Define ${\cal C}=\partial
I^{+}(p;\langle \langle M_{\mathrm{ext}}\rangle \rangle )$.  Then
${\cal C}$ automatically is an achronal, Lipschitz hypersurface. We
write ${\cal C}_{{\mathrm{ext}}}={\cal C}\cap M_{{\mathrm{ext}}}$. The
following has been proved in \cite{ChWald}:

\begin{Theorem}[P.C. \& R. Wald, 94]
\label{T1} Let $(M,g_{ab})$ be a stationary 
{spacetime}
containing a single asymptotically flat region, whose domain of outer
\communications, $\langle \langle M_{\mathrm{ext}}\rangle \rangle $,
is globally 
hyperbolic. Suppose that the null energy condition (\ref{energy}) holds.
Assume that there exists an achronal, asymptotically flat slice, ${\cal S}$,
of $\langle \langle M_{\mathrm{ext}}\rangle \rangle $, whose boundary in $M$
intersects the event horizon, ${\cal H}$, of any black holes in $M$ in a
compact cross--section, $K$. If ${\cal C}\setminus {\cal
C}_{{\mathrm{ext}}}$ has 
compact closure in M (where ${\cal C}$ and ${\cal C}_{{\mathrm{ext}}}$
were defined 
above), then each connected component of $K$ is homeomorphic to a sphere.
\end{Theorem}

Theorem \ref{T1} above assumes that $(M,g_{ab})$ contains only one
asymptotically flat region. This has been done purely for notational
convenience, as we have the following consequence of \cite
{FriedmanSchleichWitt} ({\em cf.\/} \cite[Prop. 1]{ChWald}):

\begin{Proposition}
\label{P1} Let $(M,g_{ab})$ be a globally hyperbolic space--time with Cauchy
surface $\Sigma $ and with a one parameter group of isometries, $\phi _t$,
generated by a Killing vector field $X^a$. Suppose that $\Sigma $ contains a
(possibly infinite) number of asymptotic regions $\Sigma _i$, in which $X^a$
is timelike and tends asymptotically to a non--zero multiple of the unit
normal to $\Sigma $ as the distance away from some fixed point $p\in \Sigma $
tends to infinity. Consider an asymptotically flat three--end $\Sigma _i$,
and let ${\cal B}_i$ and ${\cal W}_i$ be the black-- and white--hole regions
with respect to $\Sigma _i$ as defined above. Consider the domain of outer
\communication\  $\langle \langle M_i\rangle \rangle $ (defined as in (\ref
{mext})-(\ref{doc}), using $\Sigma _i$) %$$
%\langle\langle\rh_i\rangle\rangle =M\backslash\{\B_i\cup \W_i\}.
%$$ 
If the null energy condition (\ref{energy}) holds, %\begin{equation}
%R_{ab}Y^aY^b\geq 0 \ \mbox{ for all null }\  Y^a\ , \label{NEC}
%\end{equation}
then 
\[
M_i\cap \,J^{\pm }(\langle \langle M_j\rangle \rangle )=\emptyset \ 
\mbox{
for }\ i\neq j\,. 
\]
\end{Proposition}

This result shows, in particular, that in the Lichnerowicz--type Theorem 2.7
of \cite{Chnohair} the condition (2.11) there is unnecessary,

There is of course an equivalent result in the context of Galloway's
Theorem \ref{T:galloway}, under appropriate conditions on
$\Scri$.

\section{The structure of isometry groups of asymptotically flat
space--times.}

\label{sec:groups}

A prerequisite for studying stationary space--times is the understanding of
the structure of the isometry groups which can arise, together with their
actions. A reasonable restriction which one may wish to impose, in addition
to asymptotic flatness, is that of timelikeness of the ADM momentum of the
space--times under consideration. As is well known, this property will hold
in all space--times which are sufficiently regular and satisfy an energy
condition ({\em cf.\/} \cite{Horowitz,ChBeig1} for a list of references).
The first complete proof, in the context of black hole space--times, has
been recently given by \cite{Herzlich}. For the theorem that follows we do
not assume anything about the nature of the Killing vectors or of the matter
present; it is therefore convenient to use a notion of asymptotic flatness
which uses at the outset four--dimensional coordinates. A metric on $\Omega$
will be said to be {\em asymptotically flat} if there exist $\alpha>0$ and $
k\ge 0$ such that 
\begin{equation}  \label{D1.new}
|g_{\mu\nu} -\eta_{\mu\nu}| + r|\partial_{\alpha} g_{\mu\nu}| + \cdots +
r^k|\partial_{\alpha_1}\cdots\partial_{\alpha_k} g_{\mu\nu}| \le C
r^{-\alpha}
\end{equation}
for some constant $C$ ($\eta_{\mu\nu}$ is the Minkowski metric). $\Omega$
will be called {\em a boost--type domain}, if 
\begin{equation}
\Omega=\big\{(t,\vec x)\in {\Bbb R} \times {\Bbb R}^3: |\vec x|\ge R, |t|\le
\theta r+C\big\}\,,  \label{D.Omega}
\end{equation}
for some constants $\theta>0$ and $C\in{\Bbb R}$. Let $\phi_t$ denote
the flow of a Killing vector field X. $(M,g_{\mu\nu})$ will be said to
be {\em stationary--rotating} if the matrix of partial derivatives
of $X^\mu$ asymptotically approaches a rotation matrix in $\Sext$, and
if $\phi_t$ moreover satisfies
\[
\phi_{2\pi}(x^\mu)= x^\mu +A^\mu + O(r^{-\epsilon}), \qquad \epsilon > 0 
\]
in the asymptotically flat end, where $A^\mu$ is a timelike vector of
Minkowski space--time (in particular $A^\mu \ne 0$). One can
think of $\partial/\partial \phi + a \partial/\partial t$, $a\ne 0$ as a
model for the behavior involved. The interest of that definition stems from
the following result, proved in \cite{ChBeig2}:

\begin{Theorem}[P.C. \& R. Beig, 96]
\label{T0} Let $(M,g_{\mu \nu })$ be a space--time containing an
asymptotically flat boost--type domain $\Omega $, with time--like
(non--vanishing) ADM four momentum $p^\mu $, with fall--off exponent $\alpha
>1/2$ and differentiability index $k\ge 3$ (see eq.\ (\ref{D1.new}) below).
We shall also assume that the hypersurface $\{t=0\}\subset \Omega $ can be
Lorentz transformed to a hypersurface in $\Omega $ which is asymptotically
orthogonal to $p^\mu $. Suppose moreover that the Einstein tensor $G_{\mu
\nu }$ of $g_{\mu \nu }$ satisfies in $\Omega $ the fall--off condition 
\begin{equation}
G_{\mu \nu }=O(r^{-3-\epsilon }),\qquad \epsilon >0\ .  \label{Efalloff}
\end{equation}
Let $G_0$ denote the connected component of the group of all isometries of $
(M,g_{\mu \nu })$. If $G_0$ is non--trivial, then one of the following holds:

\begin{enumerate}
\item  $G_0={\Bbb R}$, and $(M,g_{\mu \nu })$ is either stationary, or
stationary--rotating.

\item  $G_0=U(1)$, and $(M,g_{\mu \nu })$ is axisymmetric.

\item  $G_0={\Bbb R}\times U(1)$, and $(M,g_{\mu \nu })$ is
stationary--axisymmetric.

\item  $G_0=SO(3)$, and $(M,g_{\mu \nu })$ is spherically symmetric.

\item  $G_0={\Bbb R}\times SO(3)$, and $(M,g_{\mu \nu })$ is
stationary--spherically symmetric.
\end{enumerate}
\end{Theorem}

The reader should notice that Theorem \ref{T0} excludes boost--type Killing
vectors (as well as various other behavior). This feature is specific to
asymptotic flatness at spatial infinity; see \cite{BicakSchmidt} for a large
class of vacuum space--times with boost symmetries which are asymptotically
flat in light--like directions. The theorem is sharp, in the sense that the
result is not true if $p^\mu$ is allowed to vanish or to be non--time--like.

We find it likely that there exist no electro--vacuum,
asymptotically flat space--times
which have no  black hole region, which are stationary--rotating and
for which $G_0=\R$. A similar statement should be true for domains
of outer \communications\  of regular black hole space--times. It would
be of interest to prove this result. Let us also point out that the
Jacobi ellipsoids \cite{Chandra:ellipso} provide a Newtonian example
of solutions with a one dimensional group of symmetries with a
``stationary--rotating'' behavior.

Theorem \ref{T0} is used in the proof of Theorem \ref{T:rigidity} below.

\section{Killing vectors vs. isometry groups}

\label{sec:completeness}

In general relativity there exist at least two ways for a solution to be
symmetric: there might exist

\begin{enumerate}
\item  \label{field} a Killing vector field $X$ on the space--time $(M,g)$,
or there might exist

\item  \label{globalaction} an action of a (non--trivial) connected Lie
group $G$ on $M$ by isometries.
\end{enumerate}

Clearly \ref{globalaction} implies \ref{field}, {\em but \ref{field}
  does not need to imply \ref{globalaction}} (remove {\em e.g.\/}
points from a space--time on which an action of $G$ exists). In the
uniqueness theory, as presented {\em e.g.\/} in
\cite{HE,CarterCargese,Heusler:book,Weinstein3}, one {\em always}
assumes that an action of a group $G$ on $M$ exists. This is
equivalent to the statement, that the orbits of all the (relevant)
Killing vector fields are complete. In \cite{Chorbits} and
\cite{Chnohair} completeness of orbits of Killing vectors was shown
for vacuum and electro--vacuum space--times, under various conditions.
The results obtained there are not completely satisfactory in the
black hole context, as they do not cover degenerate black holes.
Moreover, in the case of non--degenerate black holes, the theorems
proved there assume that all the horizons contain their bifurcation
surfaces, a condition which one may wish not to impose {\em a
  priori\/} in some situations ({\em cf.\/} the discussion in Section
\ref{sec:bifurcation} below). The following result, which takes care
of those problems and which does {\em not\/} assume any field
equations, has been proved in \cite{Ch:rigidity}:

\begin{Theorem}
\label{T2}
Consider a space--time $ (M,g_{ab})$ with a Killing vector field $X$
and suppose that $M$ contains an asymptotically flat three--end
$\Sigma _{\mathrm{ext}}$, with $X$  time--like in $ \Sigma
_{\mathrm{ext}}$. (Here the metric is assumed to be twice
differentiable, while asymptotic flatness is defined in the sense of
eq.\ (\ref{D1.new}) with $ \alpha >0$ and $k\ge 0$.) Suppose that the
orbits of $X$ are complete through all points $p\in \Sigma
_{\mathrm{ext}}$.
%Let $\langle\langle M_{\mathrm
%    ext}\rangle\rangle$ denote the domain of outer \communications\ 
%  associated with $\Sext$ as defined in eq. \eq{doc}. 
If $\langle \langle M_{\mathrm{ext}}\rangle \rangle $ is globally
hyperbolic, then the orbits of $X$ through points $p\in \langle
\langle M_{\mathrm{ext}}\rangle \rangle $ are complete.
\end{Theorem}

In \cite{Ch:rigidity} a generalization of this result to stationary--rotating
space--times has also been given. A theorem of Nomizu \cite{Nomizu},
together with Theorem \ref{T0} %and \ref{T2} 
give the following result \cite{Ch:rigidity}, which will be used in
the proof of Theorem \ref{T:rigidity2} below:

\begin{Theorem}
\label{T:rigidity} Consider an analytic  space--time
$(M,g_{ab})$ with a 
Killing vector field $X$ with complete orbits. Suppose that $M$
contains an asymptotically flat three--end $\Sigma _{\mathrm{ext}}$
with time--like ADM four--momentum, and with $X(p)$  time--like for
$p\in \Sigma _{\mathrm{ext}}$. (Here
asymptotic flatness is defined in the sense of eq.\ (\ref{falloff}) with $
\alpha >1/2$ and $k\ge 3$, together with eq.\ (\ref{Efalloff}) .)
 Let $\langle \langle M_{ext}\rangle \rangle $ denote the
domain of outer \communications\  associated with $\Sigma
_{\mathrm{ext}}$ as defined below; assume that $\langle \langle
M_{\mathrm{ext}}\rangle \rangle $ is globally hyperbolic and simply
connected. If there exists a Killing vector field $Y$, which is not a
constant multiple of $X$, defined on an open subset ${\cal O}$ of
$\langle \langle M_{\mathrm{ext}}\rangle \rangle $, then the isometry
group of $\langle \langle M_{\mathrm{ext}}\rangle \rangle $ (with the
metric obtained from $(M,g_{ab})$ by restriction) contains ${\Bbb
  R}\times U(1)$.
\end{Theorem}

We emphasize that no field equations or energy inequalities are assumed
above. Note that simple connectedness of the domain of outer \communications\ 
necessarily holds when a positivity condition is imposed on the Einstein
tensor of $g_{ab}$, as shown by Galloway's Theorem \ref{T:galloway} above.
Similarly the hypothesis of time--likeness of the ADM momentum will follow if
one assumes existence of an appropriate space--like surface in $(M,g_{ab})$.
It should be emphasized that no claims about isometries of $M\setminus
\langle\langle M_{\mathrm{ext}}\rangle\rangle$ (with the obvious metric) are made.

\section{The rigidity theorem}

\label{sec:rigidity}

Let us briefly recall the strategy used in \cite{HE} to prove the rigidity
theorem: If the Killing vector field $X$ is not tangent to the generators of
the horizon, the authors of \cite{HE} argue that there exists another
Killing vector $Y$ defined in a neighborhood of the horizon. On p. 329 of 
\cite{HE} they next assert: ``... One then extends the isometries by
analytic continuation. ...'' This last step does not work, the underlying
reason being essentially that {\em maximal analytic extensions are not
unique}. As we show below, Theorem \ref{T:rigidity} above takes care of this
problem, at the price of a somewhat weaker conclusion.

It is conceivable that the set of hypotheses of \cite{HE} can be modified so
that the desired conclusion, perhaps in the form of a statement concerning
isometries of the d.o.c., can be achieved. It should, however, be stressed
that the general setup of \cite{HE} does not allow Majumdar--Papapetrou
black holes, and it is likely that no degenerate black holes can fit into
this setup. On the other hand our formulation below is compatible with
geometries of Majumdar--Papapetrou type. To be precise, we have the
following theorem, the proof of which is a mixture of the arguments of \cite
{VinceJimcompactCauchyCMP} and \cite{HE}, together with Theorem
\ref{T:rigidity}: 

\begin{Theorem}
\label{T:rigidity2} Consider an analytic %stably causal 
stationary
electro--vacuum space--time $(M,g_{ab})$ with Killing vector field $X$,
and let  $\langle \langle M_{\mathrm{ext}}\rangle \rangle $ be an
asymptotically flat globally hyperbolic domain of outer \communication\ 
in $M$. (Here asymptotic flatness is understood in the sense of
(\ref{falloff}) with $\alpha >0$ and $k\ge 1$, together with eq.
\eq{Efalloff}.) Assume that there exists an achronal asymptotically
flat slice ${\cal S}$ of $\langle \langle M_{\mathrm{ext}}\rangle
\rangle $, {} whose boundary in $M$ intersects the event horizon
${\cal E}$ of the black hole region in a compact cross--section $K$.
%Suppose, next, that ${\cal C}\setminus {\cal C}_{{\mathrm{ext}}}$ has
%compact closure in M (where ${\cal C}$ and $ {\cal
%  C}_{{\mathrm{ext}}}$ were defined in Section \ref{sec:topology}
%above). 
Suppose finally that there exists a connected component of
${\cal E}\cap J^{+}(M_{\mathrm{ext}})$, say ${\cal E}_1$, which is an
analytic submanifold of $M$.  If $X$ is {\em not} tangent to the
generators of ${\cal E}_1$, then:
\begin{enumerate}
\item  The isometry group of $\langle \langle M_{\mathrm{ext}}\rangle \rangle $
contains ${\Bbb R}\times U(1)$.
\item %If ${\cal E}_1$ is two--sided, than i
The event horizon is a Killing horizon in the following sense:
There exists a neighborhood of
  $\overline{\langle \langle M_{\mathrm{ext}}\rangle \rangle }$ in
  $J^{+}(M_{\mathrm{ext}})$ on which a Killing vector field $Z$,
%  which is
tangent to the generators of ${\cal E}_1$, is defined.
 %\cap J^+(M_{\mathrm ext})$.
%\item 
%  there exists a Killing vector field $Z$ 
%on  $\langle \langle
%  M_{\mathrm{ext}}\rangle \rangle $ which can be extended
\end{enumerate}
\end{Theorem}

Let us point out that the above result generalizes immediately to
stationary--rotating space--times. The only supplementary hypothesis needed
is that of asymptotic flatness with $\alpha>1/2$ and $k\ge 3$ in the sense
of eq. (\ref{D1.new}) (in the stationary case that property follows from the
fall--off hypotheses spelled--out above).

To prove Theorem \ref{T:rigidity2}, the following result will be needed:

\begin{Lemma}
\label{L:1} Under the hypotheses of Theorem \ref{T:rigidity2} there exists a
neighborhood ${\cal O}$ of $\overline{\langle \langle
M_{\mathrm{ext}}\rangle \rangle } $ in the space--time
$(J^{+}(M_{\mathrm{ext}}), g|_{J^{+}(M_{\mathrm{ext}})})$ such that
the following hold: 
\begin{enumerate}
\item  For all $p\in {\cal O}$ we have $X(p)\ne 0$.
\item  There exists a a time function $t$ on ${\cal O}$ such that the one
parameter group of isometries $\phi _t$ generated by $X$ acts on ${\cal O}$
by time translations: 
\[
t(\phi _s(p))=t(p)+s\ . 
\]
(In particular ${\cal O}$ is invariant under $\phi _t$.)
\end{enumerate}
\end{Lemma}

Lemma \ref{L:1} can be established by a straightforward (though somewhat
lengthy) adaptation of the proofs in \cite[Section 3]{ChWald1} to the
current situation. A simplifying strategy is to construct  $t$ in two
steps, the first being the construction of $t$ on
$\overline{\langle \langle
M_{\mathrm{ext}}\rangle \rangle }\cap J^{+}(M_{\mathrm{ext}})$.

We can now pass to the proof of Theorem \ref{T:rigidity2}. By Theorem
\ref{T:galloway} and Lemma \ref{L:1} each connected component of $K$
is diffeomorphic to a 
sphere  ({\em cf.\/} the arguments in \cite{ChWald}). Let us 
choose one of those components, say $K_1$. There exists a neighborhood $
{\cal U}$ of $K_1$ in ${\cal E} _1$ together with a coordinate system $
(u,x^a=(\theta,\varphi))$ in which the tensor field $\hat g$ induced from $g$
on ${\cal U}$ takes the form 
\[
\hat g = g_{ab}dx^adx^{b} \ , 
\]
with $\frac{\partial}{\partial u }$ -- tangent to the generators of ${\cal E}
_1$. It is easily seen that $X$ is tangent to ${\cal E} $, so that on ${\cal 
U}$ the Killing vector field $X$ can be written in the form 
\[
X^\mu\partial_\mu =X^0\partial_0+ Z^a\partial_a\ . 
\]
From the Killing equation ${\cal L}_X g = 0$ we obtain 
\begin{equation}  \label{exp.1}
{\frac{\partial Z^a }{\partial u}} = 0\ , \qquad {\cal L}_Z
(g_{ab}dx^adx^{b}) = -X^0 {\frac{\partial g_{ab}}{\partial u}} dx^a dx^b\ .
\end{equation}
Let $\omega_{ab}$, $\sigma_{ab}$ and $\hat \theta$ be the vorticity, shear
and expansion of ${\cal E} _1$, as defined in \cite[p. 88]{HE}. We have $
\omega_{ab}=0$ as the generators of ${\cal E} _1$ are normal to ${\cal E} _1$
. According to Hawking and Ellis \cite[Prop. 9.31]{HE} one also has $
\hat\theta=\sigma_{ab}=0$, 
 which together with (\ref{exp.1}) implies that 
\begin{equation}  \label{Kil.1}
{\cal L}_Z (g_{ab}dx^adx^{b}) =0\ .
\end{equation}
It follows that $Z\equiv Z^a\partial_a$ is a Killing vector field on $S^2$.
By hypothesis we have $Z\ne 0$. It is well known that the orbits of every
Killing vector on $S^2$ are periodic. Let $T$ denote the (smallest positive)
period of $Z$. Let $\Psi=\phi_T$, where as before $\phi_s$ denotes the flow
of $X$. Let $G\subset {\cal \ D}\mbox{iff}({\cal O} )$ be the group
generated by $\Psi|_\cO $. Lemma \ref{L:1} shows that $\hat M\equiv {\cal O}
/G$ is a manifold. A result of Nomizu, which we present in Appendix
\ref{isometries:analyticity}, shows that $ \Psi$ is analytic, so that
$\hat M$ is an analytic manifold. Let $\Pi$ be 
the natural projection map. It follows from point 2 of Lemma \ref{L:1} that $
\hat M$ is diffeomorphic to $\Sigma\times S^1$, where $\Sigma $ is any level
set of the time function $t$ of Lemma \ref{L:1}. 
Note that our  assumption that the space--time is time--orientable (see
footnote 1)
implies that ${\cal E} _1$ is two--sided. This, in turn, shows that
${\cal S}\cap {\cal E} _1$ is two--sided in ${\cal S}$.
% Theorem \ref{T:galloway}  
The spherical topology of $K_1$ implies then that 
we can choose ${\cal O} $ so that $\Sigma$ is simply
connected. Moreover, by
construction, $\Pi {\cal E} _1$ is a hypersurface diffeomorphic to $
S^2\times S^1$ which is ruled by closed null geodesics. Theorem 1 of 
\cite[p. 404]{VinceJimcompactCauchyCMP} shows that there exists a
neighborhood ${\cal V}$ of $\Pi {\cal E} _1$ on which a Killing vector
field $\hat Z$ is defined, which is tangent to the generators of $\Pi
{\cal E} _1$. Let $\hat { {\cal V}}\subset{\cal O} $ be any open set
satisfying $\hat {{\cal V}}\cap {\cal E} _1 \ne \emptyset$ such that
$\Pi|_{\hat {{\cal V}}}$ is a diffeomorphism between $\hat {{\cal V}}$
and $\Pi\hat {{\cal V}}$, with $\Pi\hat {{\cal V}}\subset{\cal V}$. We
define $Z|_{\hat {{\cal V} }}=(\Pi|_{\hat {{\cal V}}}^{-1})_*\hat Z$;
clearly $Z|_{\hat {{\cal V}}}$ is a Killing vector field on $\hat
{{\cal V}}$ tangent to the generators of $ {\cal E} _1\cap\hat {{\cal
V}}$. The asymptotic hypotheses on $\alpha$ and $k$ needed in
Theorem~\ref{T:rigidity} are 
satisfied by \cite[Prop.~1.9]{Chnohair}, and the ADM four--momentum of
${\cal S}$ is timelike by \cite{Herzlich}. Our claims follow now by
Theorem \ref{T:rigidity}. \hfill $\Box$

The most unsatisfactory feature of the rigidity theorem is the
hypothesis of analyticity of the metric in a neighborhood of the
event horizon, for which we have no justification. In this context it
is worthwile mentioning an example of a black hole vacuum space--time
(with cosmological constant) considered recently by Bi\v{c}\'ak and
Podolsk\'y \cite{Bicak:podolsky}. In that paper Bi\v{c}\'ak and
Podolsk\'y show that there exist (real analytic) Robinson--Trautman
spacetimes which can be smoothly but {\em not} analytically extended
through an event horizon to another (real analytic) Robinson--Trautman
space--time: while the metric is smooth everywhere, it is analytic
everywhere except  on the event horizon. 

\section{Bifurcation surfaces}

\label{sec:bifurcation}

Let us start with some terminology. Recall, first, that the surface gravity
of a null hypersurface to which a Killing vector field $\zeta$ is tangent is
defined by the equation 
\[
\nabla^a(\zeta^b\zeta_b)=-2\kappa\zeta^a\ . 
\]
A black hole is called degenerate when $\kappa\equiv 0$, and non--degenerate
when $\kappa$ has no zeros.

Consider, next, the set ${\cal N}=\{X^\mu X_\mu =0,X\ne 0\}$ in the
Kruszkal--Szekeres--Schwarzschild space--time, where $X$ is the standard ``$
\partial /\partial t$'' Killing vector. ${\cal N}$ has four connected
components ${\cal N}_a$, $a=1,\ldots ,4$, and for each $a$ the set $
\overline{{\cal N}_a}\setminus {\cal N}_a$ consists of the same two--sphere,
namely the set of points ${\cal S}$ where $X$ vanishes. ${\cal S}$ is called
the bifurcation surface of the bifurcate horizon ${\cal N}\cup {\cal S}$.
Whenever we have a non--degenerate Killing horizon, it is extremely
convenient for technical reasons to have the property that this horizon
comprises a compact bifurcation surface. For example, this hypothesis is
made throughout the classification theory of static (non--degenerate) black
holes ({\em cf.\ e.g.\/}
\cite{bunting:masood,Sudarsky:wald,Chnohair}). 
The problem is, that while we have  good control of the
geometry of the domain of outer \communications, various unpleasant
things can happen at its boundary. In particular, in 
\cite{RaczWald2} it has been shown that there might be an obstruction for the
extendability of a domain of outer \communication\  in such a way that
the extension comprises a compact bifuraction surface. Nevertheless,
as far as applications are concerned, it suffices to have the
following: Given a space--time $(M,g)$ with a domain of outer
\communication\  $\langle \langle M_{\mathrm{ext}}\rangle \rangle $ and a
non--degenerate Killing horizon, there exists a space--time
$(M^{\prime },g^{\prime })$, with a domain of outer \communications\ 
$\langle \langle M_{\mathrm{ext}}^{\prime }\rangle \rangle $ which is
isometrically diffeomorphic to $
\langle \langle M_{\mathrm{ext}}\rangle \rangle $, such that all
non--degenerate 
Killing horizons in $(M^{\prime },g^{\prime })$ contain their bifurcation
surfaces. R\'{a}cz and Wald have shown \cite{RaczWald2}, under
appropriate conditions, that 
this is indeed the case:

\begin{Theorem}[I. R\'acz \& R. Wald, 96]
\label{T:bifurcate} Let $(M,g_{ab})$ be a stationary, or stationary--rotating
space--time with Killing vector field $X$ and with an asymptotically
flat region $M_{\mathrm{ext}}$. Suppose that $J^{+}(M_{\mathrm{ext}})$
is globally hyperbolic with asymptotically flat Cauchy surface $\Sigma
$ which intersects the event horizon ${\cal N}=\partial \langle
\langle M_{\mathrm{ext}}\rangle \rangle \cap J^{+}(M_{\mathrm{ext}})$
in a compact cross--section. Suppose that $X$ is tangent to the
generators of ${\cal N}$ and that the surface gravity of every
connected component of ${\cal N}$ is a non--zero constant. Then there
exists a space--time $(M^{\prime },g_{ab}^{\prime })$ and an isometric
embedding
\[
\Psi :\langle \langle M_{\mathrm{ext}}\rangle \rangle \to \langle \langle
M_{\mathrm{ext}}^{\prime }\rangle \rangle \subset M^{\prime }\ , 
\]
where $\langle \langle M_{\mathrm{ext}}^{\prime }\rangle \rangle $ is
a domain of outer \communication\  in $M^{\prime }$, such that:

\begin{enumerate}
\item  {} There exists a one--parameter group of isometries of $(M^{\prime
},g_{ab}^{\prime })$, such that the associated Killing vector field $
X^{\prime }$ coincides with $\Psi ^{*}X$ on $\langle \langle
M_{\mathrm{ext}}^{\prime }\rangle \rangle $.

\item  Every connected component of $\partial \langle \langle
M_{\mathrm{ext}}^{\prime 
}\rangle \rangle $ is a Killing horizon which comprises a compact
bifurcation surface.

\item  {} There exists a ``wedge--reflection'' isometry about every connected
component of the bifurcation surface.
\end{enumerate}
\end{Theorem}

It should be emphasized that neither field equations, nor energy
inequalities, nor  
analyticity have been assumed above. In \cite{RaczWald2} conditions
are given which 
guarantee that $\kappa$ is constant on every connected component of
a Killing horizon (recall that this property holds for
{\em e.g.\/} electro--vacuum black holes).

It might be worthwile to mention that the above conclusion about
existence of ``wedge--reflection'' isometries has been established in
\cite{RaczWald2} for connected event horizons only. This claim can,
however, easily be generalized to the multiple black hole case using
an obvious extension of the construction in \cite{RaczWald2},
together with the Kuratowski--Zorn lemma. 

\section{Uniqueness of static, degenerate black holes}

\label{sec:degenerate}

An important part of the classification program of stationary
space--times is the classification of static black holes. Until
recently the existing theory covered only non--degenerate black
holes ({\em
  cf.}~\cite{Israel:uniqueness,CarterCargese,Ruback,bunting:masood,%
  Sudarsky:wald,Chnohair}).  For the sake of completeness it is
clearly of interest to include the degenerate ones in the
classification. Recall, moreover, that degenerate black holes play
some role in quantum theories \cite{Gibbons:inBarrow}, {\em e.g.\/} in
string theory ({\em cf.\ e.g.\/}~\cite{Garyonbh}). It is widely expected
that the only such space--times are the ``standard Majumdar--Papapetrou
black holes''.
More precisely, consider a metric $g$ and an electro--magnetic potential $A$
of the form \cite{Majumdar,Papapetrou:MP} 
\begin{eqnarray}
&g=-u^{-2}dt^2+u^2(dx^2+dy^2+dz^2)\,,&  \label{I.0} \\
&A=u^{-1}dt\,,& \label{I.0.1}
\end{eqnarray}
with some nowhere vanishing, say positive, function $u$. Einstein--Maxwell
equations read then 
\begin{equation}
\frac{\partial u}{\partial t}=0\,,\qquad \frac{\partial ^2u}{\partial x^2}+
\frac{\partial ^2u}{\partial y^2}+\frac{\partial ^2u}{\partial z^2}=0\,.
\label{I.1}
\end{equation}
A space--time will be called a standard MP space--time if the coordinates $
x^\mu $ of (\ref{I.0})--(\ref{I.0.1}) cover the range ${\Bbb R}\times ({\Bbb 
R}^3\setminus \{\vec{a}_i\})$ for a finite set of points $\vec{a}_i\in {\Bbb 
R}^3$, $i=1,\ldots ,I$, and if the function $u$ has the form 
\begin{equation}
u=1+\sum_{i=1}^I\frac{m_i}{|\vec{x}-\vec{a}_i|}\,,  \label{standard}
\end{equation}
for some positive constants $m_i$. It has been shown by Hartle and Hawking 
\cite{HartleHawking} that every standard MP space--time can be analytically
extended to an electro--vacuum space--time with a non--empty black hole
region, and with a domain of outer \communication\  which is non--singular in
the sense described in Theorem \ref{T:mp} below. It is of interest to
enquire whether the standard Majumdar--Papapetrou metrics describe all
possible regular black holes with a metric of the form (\ref{I.0}). For the
purpose of classification of black holes it is actually necessary to allow
metrics of the form (\ref{I.0}) for which the coordinates are not
necessarily global ones: we shall say that a space--time $(M,g)$ is locally
a MP space--time if for every point $p$ there exists a coordinate system
defined in a neighborhood of this point such that (\ref{I.0})--(\ref{I.0.1}%
) holds. 
We have 
the following result of M. Heusler \cite{heuslerMP}:

\begin{Theorem}[M. Heusler, 96]\label{T:Heusler}
  Let $(M,g)$ be a static  electro--vacuum space--time,
% with a simply  connected domain of outer \communication. S
  and suppose that there exists in $M$ an asymptotically flat simply
  connected slice ${\cal S}\subset \langle \langle
  M_{\mathrm{ext}}\rangle \rangle $ with compact interior and compact
  boundary $\partial {\cal S}\subset {\cal E}=\partial \langle \langle
  M_{\mathrm{ext}}\rangle \rangle $ $\cap J^{+}(M_{\mathrm{ext}}).$
  (Here asymptotic flatness is defined in the sense of eq.\ 
  (\ref{falloff}) with $ \alpha >0$ and $k\ge 1$, together with eq.\ 
  (\ref{Efalloff}).)  For every connected component $\partial {\cal
    S}_{a}$ of $\partial {\cal S}$ we set
\[
Q_a=-\frac 1{4\pi }\int_{\partial {\cal S}_{{a}}}*F,
\]
where $F$ is the electro--magnetic field two--form. If all the components of $
{\cal E}$ are degenerate, and if 
%all the charges $Q_a$ are of the same sign, 
$Q_a Q_b \ge 0$ for all pairs of indices $a,b$, then $(M,g)$ is
locally a MP space--time. 
\end{Theorem}

It would be of interest to exclude the possibility of existence of
non--connected static black holes with degenerate horizons and with
$Q_aQ_b<0$ for some pair of indices $a$ and $b$. The hypothesis of
staticity can be replaced by that of stationarity if one further
assumes that the event horizon is connected, that the global angular
momentum of $\Sext$ vanishes and that there are no ergo--regions, see
\cite{Heusler:book}.

In the case of a connected black hole the idea of proof of Theorem
\ref{T:Heusler} is rather simple. In such a case the following
mass--squared identity has been established in \cite{Heusler:book},
see also \cite{heuslerMP} and Section~12\ of the accompanying lectures
of Heusler \cite{Heusler:ascona}:
\begin{equation}
\label{massidentity}
Q^2+\left( \frac{\kappa A}{4\pi }\right) ^2 =M^2, 
\end{equation}
where A is the area of $\partial {\cal S}$. To prove this identity one
can use Carter's result that there are no ergo--regions in static domains
of outer \communication, together with simple connectedness of the
d.o.c. Let us mention that the above identity has already been shown
by Israel under the hypothesis that the horizon comprises its
bifurcation surface (in particular it cannot be degenerate). Eq.
\eq{massidentity} gives $|Q|=M$ when $\kappa $ vanishes, which implies
an identity which, together with an
observation of Israel and Wilson shows that the metric is, locally, a {\em 
Majumdar--Papapetrou} metric.

A classification theorem for a class of static degenerate black holes
will follow from the 
above and from the following result \cite{ChNad}:

\begin{Theorem}[P.C. \& N. Nadirashvili, 95]
\label{T:mp} Consider an electro--vacuum space--time $(M,g)$ with a
non--empty black hole region ${\cal B}$. Suppose that there exists in
$M$ an asymptotically flat space--like hypersurface $\Sigma $ with
compact interior and with boundary $\partial \Sigma \subset {\cal B}$.
(Here asymptotic flatness is defined in the sense of eq.\ 
(\ref{falloff}) with $ \alpha >0$ and $k\ge 1$, together with eq.\ 
(\ref{Efalloff}).)  
 %and with timelike (non--vanishing) ADM four--momentum. 
Assume moreover that on the closure of the domain of outer \communication\  $
\langle \langle M_{\mathrm{ext}}\rangle \rangle $ there exists a
Killing vector field $ X$ with complete orbits diffeomorphic to ${\Bbb
  R}$, $X$ being timelike in an asymptotic region of $\Sigma $. If
$(M,g)$ is locally a MP space--time in the sense described above, then
$\langle \langle M_{\mathrm{ext}}\rangle \rangle $ is isometrically
diffeomorphic to a standard MP space--time.
\end{Theorem}

{\bf Remark:} Let us mention that, in particular, under the hypotheses
above, every connected component of the event horizon has to carry
charge, and all charges have to be of the same sign.

\proof By \cite{Herzlich} the ADM mass of $\Sigma $ is timelike, so
that Theorem 1 of \cite{ChNad} applies ({\em cf. }also Appendix
\ref{app:mp} where one of the steps of the proof of that result is
justified in more detail than in \cite{ChNad}). It remains to show
that the set $\hat{M}$ defined in \cite {ChNad} coincides with
$\langle \langle M_{\mathrm{ext}}\rangle \rangle $. This will be
established by arguments rather similar to those used in Section 3 of
\cite{ChWald1}. By \cite{ChNad} we have $\hat{M}$ $\subset \langle
\langle M_{\mathrm{ext}}\rangle \rangle .$ Suppose, for contradiction,
that $\partial \hat{M}$ $\cap \langle \langle M_{\mathrm{ext}}\rangle
\rangle \neq \emptyset$, thus there exists a point $p\in \partial
\hat{M}$ with $p\in \langle \langle M_{\mathrm{ext}}\rangle \rangle $.
Let $r$ be a point in $\langle \langle M_{\mathrm{ext}}\rangle \rangle
$ such that there exists a causal future directed curve $\gamma $ from
$r$ to $p$. We wish to show that there exists $T$ such that $\phi
_T(p)\in I^{+}(\Sigma )$. Indeed, if $r$ $\in I^{+}(\Sigma )$ we
are done, otherwise there exists $T$ such that $\phi _T(r)\in I^{+}(\Sigma )$
, and our claim follows as $\phi _T(p)\in \phi _T(I^{+}(r))=I^{+}(\phi
_T(r))\subset I^{+}(\Sigma )$. Now $\hat{M}$ is $\phi _T$ invariant by
its definition, so is therefore $\partial \hat{M}$, so that $\phi
_T(p)\in \partial \hat{M}\cap I^{+}(\Sigma )$. We note the following:

\begin{Lemma}
\label{L:mp2}
The standard MP space--times ($M,g_{\mu \nu })$ are globally
hyperbolic, with the slices $t=const$ being Cauchy surfaces.
\end{Lemma}

\proof Let $\gamma $ be a causal curve in $M$. We can parameterize $\gamma $
by $t$, so that $\gamma (t)=(t,\vec{x}(t))$. The condition of
causality gives
\begin{equation}|\frac{d\vec{x}}{dt}|_\delta \leq u^{-2}
\label{vineq}\ ,\end{equation}
where $|.|_\delta $ denotes the norm with respect to the flat metric $\delta
_{ij}$. Let $\left\{ \vec{a}_i\right\} _{i=1}^N$ be the singular
set of $\varphi $. The hypersurfaces $t=const$ would fail to be Cauchy in $M$
if and only if $|\vec{x}(t)|$ would go to infinity in finite
time, or if $\vec{x}(t)$ would reach one of the $\vec{a%
}_i$'s in finite time. The former can be immediately excluded by the
asymptotic flatness of the metric, to exclude the latter note
that we have $\varphi \geq \frac{m_i}{r_i}$, where $r_i=$ $|
\vec{x}-\vec{a_i}|$. Eq. (\ref{vineq}) thus gives
\[
\frac{dr_i}{ds}\leq \frac{r_i^2}{m_i^2}. %\label{vineq2} 
\]
This last inequality implies
\[
|\frac 1{r_i}(t_1)-\frac 1{r_i}(t_2)|\leq \frac{|t_1-t_2|}{m_i^2}, 
\]
\[
\]
and the result follows.\hfill $\Box$

Returning to the proof of Theorem \ref{T:mp}, let $\gamma $ be a future
directed causal curve from $p$ to $M_{\mathrm{ext}}$. By Lemma \ref{L:mp2} $\gamma $
intersects every level set of $t$. Consider the asymptotically flat
hypersurface $\Sigma \cap \hat{M}$. By \cite{ChmassCMP} $\Sigma \cap 
\hat{M}$ is a graph of a function $u$, over a subset $\Omega $ of $
\left\{ t=0\right\} $, such that $u$ approaches a constant as $r$ tends to $
\infty $. Moreover $\Omega $ contains the complement of some ball. As $
\Sigma \cap \hat{M}$ is spacelike $\Omega $ is open. By the interior
compactness property of $\Sigma $ it follows that $\Omega $ is closed, hence 
$\Omega =\left\{ t=0\right\} $. Consider any future directed causal curve $
\gamma $ in $M$ such that $r$ tends to infinity on $\gamma $ towards the
future. Clearly $\varphi _t(\Sigma \cap \hat{M})$ intersects $\gamma $
for $t$ large enough. As $\gamma $ is timelike the set of $t$ such that $
\gamma $ intersects $\varphi _t(\Sigma \cap \hat{M})$ is open. By the
interior compactness property of $\Sigma $ together with global
hyperbolicity of $\hat{M}$ it is closed. It follows that $\gamma $
intersects $\Sigma \cap \hat{M}$ at some point $q$. We then have $q\in
I^{+}(p)\subset I^{+}(\Sigma )$ which contradicts achronality of $\Sigma $,
and Theorem \ref{T:mp} follows. \hfill $\Box$

Theorems \ref{T:galloway}, \ref{T:Heusler} and \ref{T:mp}  (together
with some standard arguments which we shall not reproduce here) give
the following result:

\begin{Theorem}\label{T:mp2}
  Consider a {\em static} electro--vacuum black--hole space--time
  $(M,g)$ with a globally hyperbolic domain of outer \communication.
  Suppose that $M$ contains an asymptotically flat achronal
  hypersurface $\Sigma $ which has compact interior and a compact {\em
    connected} boundary $\partial \Sigma $ located on the event
  horizon. (Here asymptotic flatness is defined in the sense of eq.\ 
  (\ref{falloff}) with $ \alpha >0$ and $k\ge 1$, together with eq.\ 
  (\ref{Efalloff}).) If the event horizon is {\em degenerate}, then
  the domain of outer \communication\  is isometrically diffeomorphic to
  that of an extreme Reissner--Nordstr\"{o}m black hole.
\end{Theorem}

\section{Uniqueness of stationary, axisymmetric, non--degenerate black holes}

\label{sec:axisymmetric}

Another key ingredient of the classification of regular black holes is
that of classification of the stationary and axisymmetric black holes.
It is known that the Kerr--Newman black holes exhaust the family of
{ stationary--axisymmetric, connected, non--degenerate, and
  appropriately regular} electro--vacuum black holes
\cite{CarterCMP,Mazur}
satisfying 
\begin{equation}
\label{acondition}
M^2> Q^2+a^2\ .
\end{equation}
Here $M$ is the total ADM mass of the black hole, $Q$ its total
electric charge and $aM$ its total angular momentum.  Some new results
concerning the non--connected case have been recently obtained by
Weinstein \cite{Weinstein3} ({\em cf.\/} also
\cite{Weinstein1,Weinstein2,Weinstein:trans}), let us discuss those
shortly.

Consider thus a stationary--axisymmetric electro--vacuum black hole
space--time, and suppose that its domain of outer \communication\  $\doc$
contains an asymptotically flat space--like hypersurface $\Sigma$ with
compact interior and compact boundary $\partial\Sigma_a \subset {\cal
  E}\equiv \partial \doc \cap J^+(\Mext)$.  Let $\partial\Sigma_a$,
$a=1,\ldots,N$ be the connected components of $\partial \Sigma$.
Let $\tau=g_{\mu\nu}X^\mu dx^\nu$, where $X^\mu$ is the Killing
vector field which asymptotically approaches the unit normal to
$\Sext$. Similarly set $\zeta=g_{\mu\nu} Y^\mu dx^\nu$, $Y^\mu$ being
the Killing vector field associated with rotations. Define
\begin{eqnarray}
& q_a=-\frac{1}{4\pi}\int_{\partial\Sigma_a}*F\ , &
\label{qa}
\\
& m_a =-\frac{1}{8\pi}\int_{\partial\Sigma_a}*d\tau\ , &
\label{ma}
%\begin{equation}
\\ 
\label{lint}
& L_a=-\frac{1}{4\pi}\int_{\partial\Sigma_a}*d\zeta\ . &
%\end{equation}
\end{eqnarray}
Here $F$ is the electro--magnetic field two--form.  Note that \eq{ma} is
the standard Komar integral associated with $X^\mu$, and $L_a$
is the Komar integral associated with the Killing vector $Y^\mu$.
It is therefore natural to think of $L_a$ as the angular
momentum of each connected component of the black hole. 
Set
\begin{equation}
  \label{mua} 
\mu_a=m_a -2 \omega_a L_a\ , 
  \end{equation}
  where $\omega_a$ is the rotation velocity of the $a$'th black hole.
  Weinstein shows that one necessarily has $\mu_a >0$.  Let $r_a >0$,
  $a=1,\ldots,N-1$, be the distance along the axis between
  neighboring black holes as measured with respect to an auxiliary
  (unphysical) metric, {\em cf.\/} \cite{Weinstein3} for details.  Let
  finally $\lambda_a$, $a=1,\ldots,N$ be the constants introduced in
  \cite{Weinstein3}. The definition of the $\lambda_a$'s involves
  various auxiliary potentials, let us simply note that in vacuum we
  have $\lambda_a=L_a$.  We have the following result of Weinstein
  \cite{Weinstein3}:
\begin{Theorem}[G. Weinstein, 96]
  \label{T:Weinstein}
  Consider a stationary--axisymmetric electro--vacuum space--time
  $(M,g)$ with a globally hyperbolic domain of outer \communications\ 
  $\doc$. Suppose that $M$ contains an asymptotically flat
  hypersurface $\Sigma$ with compact interior and compact boundary
  $\partial \Sigma=\cup_{i=1}^N\partial \Sigma_a$, where each of the
  of the $N$ connected components $\partial\Sigma_a$ of $
  \partial\Sigma$ satisfies $\partial\Sigma_a\subset {\cal E}\equiv
  \partial \doc\cap J^+(\Mext)$. (Here asymptotic flatness is defined
  in the sense of eq.\ (\ref{falloff}) with $ \alpha >0$ and $k\ge 1$,
  together with eq.\ (\ref{Efalloff}).) Suppose moreover that every
  connected component of ${\cal E}$ is non--degenerate. Then:
\begin{enumerate}
\item The inequality \eq{acondition} holds.
\item{} The metric on $\doc$ is uniquely determined (up to isometry)
  by the $4N-1$ parameters
  \begin{equation}
    \label{par}
    (\mu_1,\ldots,\mu_N,\lambda_1,\ldots,\lambda_N,q_1,\ldots,q_N,r_1,\ldots,
r_{N-1})
  \end{equation}
described above, with $r_a,\mu_a>0$.
\end{enumerate}
\end{Theorem}

Let us note that for connected black holes point 1 of Theorem
\ref{T:Weinstein} removes the condition \eq{acondition} from Mazur's
theorem \cite{Mazur}. The fact that \eq{acondition} is not necessary has
been known to some authors \cite{Fletcher,Carter:unpublished}, but we
are not aware of any previous published proof.

It is known that for some sets of parameters \eq{par} the solutions will
have ``strut singularities'' between some pairs of neighboring black
holes \cite{Weinstein:trans,Li:Tian2}. In the statement of Theorem
\ref{T:Weinstein} we have for 
simplicity assumed smoothness of the domain of outer
\communication. The conclusion of Theorem \ref{T:Weinstein} remains
valid  when strut singularities are allowed in the metric.

As far as existence of {\em vacuum} non--connected rotating black holes is
concerned, we have the following result, also due to Weinstein
\cite{Weinstein:trans}:
\begin{Theorem}[G. Weinstein, 94]
  \label{T:Weinstein2} For every $N\ge 2$ and for every set of
parameters \begin{equation} \label{par1}
(\mu_1,\ldots,\mu_N,L_1,\ldots,L_N,r_1,\ldots, r_{N-1})\ , \end{equation}
with $\mu_a,r_a>0$, there exists a vacuum space--time $(M,g)$ satisfying
the hypotheses of Theorem \ref{T:Weinstein}, except perhaps for
``strut singularities'' on the axis between some neighboring black
holes.
\end{Theorem}

The existence of the ``struts'' for all sets of parameters as above
is not known, and is the main open problem in our understanding of
stationary--axisymmetric electro--vacuum black holes.

For the electro--vacuum case, an equivalent of Theorem
\ref{T:Weinstein2} is true \cite{Weinstein3} modulo a ``regularity''
result on the singular set for the harmonic maps with prescribed
singularities into the complex hyperbolic plane, which has not been
established yet ({\em cf.\/} \cite{LiTian:regularity} for some
related results). More precisely, Weinstein has shown
\cite{Weinstein3} that for any set of parameters \eq{par} with
$\mu_a,r_a>0$ there exists a space--time $(M,g)$ satisfying the
hypotheses of Theorem \ref{T:Weinstein}, except perhaps for
singularities on the axis of rotation. One expects that those
singularities, if any, will again be ``strut singularities'' between
some neighboring black holes, but no rigorous proof of this fact is
available so far. It would be of interest to fill this gap.

\section{Differentiability of event horizons}

\label{sec:differentiability}

One of the hypotheses of Theorem \ref{T:rigidity2}, in addition to
analyticity of space--time and of the metric, is that of analyticity of the
event horizon. This is a logically independent requirement, as is clearly
demonstrated by the following result \cite{ChGalloway}:

\begin{Theorem}[P.C. \& G. Galloway, 96]
\label{T:nondifferentiability} There exist vacuum analytic space--times with
a non--empty black hole region and with a {\em nowhere differentiable}
event horizon ${\cal E}$, in the sense that no open subset of ${\cal E}$ is
a differentiable manifold.
\end{Theorem}

The example constructed in \cite{ChGalloway} is admittedly artificial, as it
is obtained by removing an appropriate subset out from Minkowski
space--time. There are certainly several desirable and independent sets of
supplementary hypotheses which one could impose to exclude this kind of
space--times. It is nevertheless worrisome, because it shows that high
differentiability of global constructs such as event horizons, Cauchy
horizons, etc., should not be taken for granted and should be justified in
situations under consideration. In particular, Theorem \ref
{T:nondifferentiability} leads immediately to the question, whether the area
theorem hold for event horizons with low differentiability. Because of the
widely accepted relationship of the area theorem with the laws of
thermodynamics, it would be of interest to analyze this problem.

\section{Conclusions}

\label{sec:conclusions}

We have discussed various  recent developments in our understanding of
space--time with black holes. From what has been said it should be clear
that several important questions remain open. We shall end this paper with
two lists of problems which, we believe, are worthwhile being investigated.
The main problems are:

\begin{enumerate}
\item  Prove the Rigidity Theorem \ref{T:rigidity2} without the hypothesis
of analyticity of the space--time and of the event horizon, or construct a
counterexample.

\item  Show that a sufficiently regular multi black--hole space--time must
be a Majumdar--Papapetrou space--time, or construct a counterexample.
\end{enumerate}

The following problems below would clearly be superseded by solutions of
their counterparts above. Nevertheless one could try to solve the problems
below, as a step towards solving the more difficult ones:

\begin{enumerate}
\item  Remove the hypothesis of analyticity of the event horizon in the
Rigidity Theorem \ref{T:rigidity2}.

\item  Show that Weinstein's non--connected (stationary, axi--symmetric,
electro--vacuum) black holes are singular ({\em cf.\/}
\cite{Weinstein:trans,Li:Tian2} for some partial results).

\item  Show that the only sufficiently regular Israel--Wilson--Perjes black
holes are in the Majumdar--Papapetrou class ({\em cf.\/} \cite
{HartleHawking,Whitt} for some partial results).

\item  Extend Weinstein's Classification Theorem \ref{T:Weinstein} to
include degenerate components of the event horizon.

\end{enumerate}

{\bf Acknowledgments} The author acknowledges useful discussions with,
and comments from, G.~Galloway. He is grateful to M.~Heusler, J.~Isenberg,
I.~R\'acz, R.~Wald and 
G.~Weinstein for comments about a previous version of this
paper. Finally he wishes to thank K. Nomizu for communicating him Theorem
\ref{T:Nomizu}, together with its proof, and for allowing him to
reproduce his proof in this paper. This work was supported in part by
the KBN grant 2P30105007.

\appendix

\section{Analyticity of isometries}
\label{isometries:analyticity}

We have the following unpublished result of Nomizu,
 which we prove here for completeness. The analyticity of isometries
 of analytic manifolds is a direct consequence of this result:

\begin{Theorem}
\label{T:Nomizu}
 Suppose $M$ and $M'$ are real analytic manifolds each
provided with a real analytic affine connection.  Then
a (smooth) diffeomorphism $f$ from $M$ onto $M'$ which
preserves the affine connections is real analytic.
\end{Theorem}
 \proof To prove the result, it suffices to note that for $M$ with an
analytic affine connection $\nabla$, the exponential mapping
$\exp_p:T_pM\rightarrow M$ is real analytic on a neighborhood of $0$ in
$T_pM$, say $U$ (so that $\exp_p(U)$ is a normal neighborhood of
$p$). We have for each $p\in M$
  $$f(\exp_p(X))=\exp_{f(p)}(AX) \quad \mbox{for every}\  X\in V\ ,$$
  where $A$ is the differential $f_*(p)$ of $f$ at $p$  and $V$
  is an open neighborhood of $0$ in $T_pM$. The analyticity of $f$
  follows immediately from this equation.
\hfill $\Box$

\section{A comment to the classification theorem of non--singular MP black
holes.}

\label{app:mp}

In \cite{ChNad}, in the paragraph following eq. (12) it was claimed
without justification that the functions $x^a $ defined there
provide a global coordinate system on the manifold $\hat{M}$ defined
there. While this claim is correct, some work is needed to justify
that. We shall present here the missing elements of the proof. We use
the notation and the definitions of  \cite{ChNad}. Let us denote by
$\Psi $ the map from $\hat{M}$ to  $
{\Bbb R}^4$ defined by the functions $x^a $. A standard asymptotic
analysis of the equations (11)--(12) of \cite{ChNad} shows that $\Psi $
is a diffeomorphism between $\Sigma _{\mathrm{ext}}\times {\Bbb R}$
and $({\Bbb R}^3\setminus K)\times {\Bbb R}$, for some compact set
$K$. Here one might need to replace the constant $R$ which defines
$\Sigma _{\mathrm{ext}}$ by a larger constant.

Consider an affinely parameterized geodesic  $\gamma $ of the metric $h_{\mu
\nu }$ defined by eq. (10) of \cite{ChNad}. From the definition of the $
x^a $'s together with the covariant constancy of the $e_\nu ^a$'s we
have
\begin{eqnarray*}
\frac{dx^a (\gamma (s))}{ds} &=&e_\nu ^a (\gamma (s))\frac{d\gamma ^\nu
(s)}{ds}\Longrightarrow \frac{D^2x^a (\gamma (s))}{ds^2}=0, \\
&&
\end{eqnarray*}
which shows that
\begin{equation}
x^a (\gamma (s))=e_\nu ^a (\gamma (s_0))\frac{d\gamma ^\nu (s_0)}{ds}
s+x^a (\gamma (s_0)).\label{linear}
\eeq
We have the following lemma:
\begin{Lemma}
\label{L:mpx} Let $\gamma :[a,b)\rightarrow \hat{M}$ be an
affinely parametrized geodesic of the metric $h_{\mu \nu }$ defined by eq.
(10) of \cite{ChNad}. If there exists $\varepsilon >0$ such that $
u|_\gamma>\varepsilon $, then $\gamma $ can be extended in $\hat{M}$ to
a geodesic defined on an interval $[a,b+\delta )$, for some $\delta >0$.
\end{Lemma}

\proof If $\overline{\gamma ([a,b))}\subset \mbox{int}(\Sigma
  _{\mathrm{ext}})\times {\Bbb 
  R}$ the result is obvious by properties of geodesics in Minkowski
space--time together with the fact, mentioned above, that $\Psi $ is a
diffeomorphism in the asymptotic region.  Note that the geodesic
$\gamma$ cannot approach the boundary of $\hat{M}$, at which $u$
vanishes, we may thus assume that $\gamma ([a,b))\subset \Sigma
_c\times {\Bbb R}$, where $\Sigma _c$ is compact. Because $t$ is a linear
function of $s$ on $\gamma $ (see eq. (\ref{linear})), we have in fact
$\gamma ([a,b))\subset \Sigma _c\times [t(\gamma (a)),t(\gamma
(a))+C(b-a)]$, for some constant $C$, which is a
compact set. Let $s_i\in [a,b)$ be any sequence such that $s_i\rightarrow b$%
, let $p$ be an accumulation point of $\gamma (s_i)$. Let $(\tau ,y^i)$ be
local MP coordinates defined in a neighbourhood of $p$, in those coordinates 
$\gamma $ is a straight line which accumulates at $p$ and clearly can be
extended as claimed. \hfill $\Box$

Consider the exponential map $\exp_q$ of the metric $h$,
$$\exp_q:T_q\hat M\supset {\cal O}_q\to \hat M$$ defined on ${\cal
  O}_q$. Lemma \ref{L:mpx} shows that ${\cal O}_q$ is of the form
$\R\times{\cal U}_q$ for some open star--shaped set ${\cal U}_q\subset
\R^3$. Moreover every maximally extended affinely parametrized geodesic
in $\hat M$ starting at $q$ is of the form $\{\exp_q(\alpha r,r\vec
n)|r\in [0,r_{\vec n})\}$ for some $ \alpha\in\R$, with either
$r_{\vec n}=\infty$ or $\lim\inf_{r\to r_{\vec n}}u(\exp_q(\alpha
r,r\vec n))=0$. Let us write a vector field $Y\in T_q\hat M$ as a
linear combination of the basis vectors $e^{a\mu}(q)$, as defined in
\cite{ChNad}, $Y=Y_a e^{a\mu}(q)$. It follows from \eq{linear} that
\begin{equation}
\label{xeq}
x^a\circ(\exp_q(Y_b e^{b\mu}(q))=Y_a + x^a(q)\ .
\end{equation}
In other words,
\beq
\label{almostinjectivity}
\Psi\circ\exp_q:{\cal O}_q\to  \R^4 \quad\mbox{is a translation. }
\eeq
In particular $\exp_q$ is injective for every $q\in\hat M$.
%For every $q\in\hat M$ w
We set 
$$\phi_q=u\circ\exp_q:{\cal O}_q\to \R\ .$$ As discussed in
\cite{ChNad} we have $\partial \phi_q/\partial t=0$ so that, by a
slight abuse of notation, we can consider $ \phi_q $ as being defined
on ${\cal U}_q$.  Moreover, again as discussed in \cite{ChNad},
$\phi_q$ satisfies
\begin{eqnarray}
&  \left(\frac{\partial^2}{\partial x^2} + \frac{\partial^2}{\partial
    y^2} +  \frac{\partial^2}{\partial z^2}\right)\phi_q =0,  & \label{ueq}
\\ 
&
|\mbox{grad}\, \phi_q^{-1}| \le C_1\ . &  \label{ueq.1}
\end{eqnarray}
Here both the gradient and its norm are taken with respect to the flat
metric on $T_q\hat M $. We have the following version of Prop.~2 of
\cite{ChNad}, the proof of which can be obtained by  trivial
modifications of the proof given in \cite{ChNad}:
\begin{Proposition}
  \label{P:ChNadmod}
Let ${\cal U}_q$ be an open subset of $\R^3$ and let $\phi_q$ be a
function satisfying \eq{ueq}--\eq{ueq.1} on ${\cal U}_q$. (It follows from
\eq{ueq.1} that $\phi_q^{-1}$ can be extended by continuity to a
function $f$ on $\overline{{\cal U}_q} $.) Set
$$
{\cal S}_q=\{p\in \overline{{\cal U}_q}\setminus {\cal U}_q| f(p)=0\}\ .
$$
Then ${\cal S}_q$ is discrete.
\end{Proposition}
Proposition \ref{P:ChNadmod} and what has been said show that for
every point $q\in \hat M $ the set ${\cal U}_q$ is $\R^3$ minus a
finite set of half--rays, ${\cal U}_q=\R^3\setminus
(\cup_{i=1}^N\{r\vec a_{i,q}, r\in[1,\infty)\}$, with $\vec a_{i,q}\ne
0$. 

We wish to show that $\Psi(\hat M)\supset \Omega\equiv \R^3\setminus
(\cup_{i=1}^N\{\vec a_{i}\})$, with $\vec a_{i}\equiv\vec a_{i,p}$.
Here $p$ is the point in $\hat M$ such that $x^a(p)=0$; we know
already that $\Omega\supset\R^3\setminus (\cup_{i=1}^N\{r\vec a_{i},
r\in[1,\infty)\})$. Consider a point $(Y_a)=(0,r\vec a_{i})$ with $r>
1$, for some $i$. We can find $r\in \Sext$ such that the triangle
$T\subset \R^4$ with vertices $0$, $x^a(r)$ and $Y_a$ satisfies
$(T\setminus \{s\vec a_{i}, s\in[1,\infty)\})\cap \partial (\R\times
{\cal O}_p)=\emptyset$.  By a simple continuity argument it can be
seen that $Y_i-x^i(r)\in {\cal U}_r$, so that by eq. \eq{xeq} we
obtain $x^a(\exp_{r}((Y_a-x^a)e^{a\mu}(r)) = Y_a$, except perhaps in
the case in which there exists a point $\vec a \in {\cal S}_p$ such
that $\vec a= C\vec a_{i,p} $, for some constant $C>1$. That last
possibility can be easily gotten rid of by replacing $p$ by a nearby
point $p'$ in the construction of the map $\Psi$.  It follows that
$\Omega\subset \Psi(\hat M)$.

For $Y\in {\cal O}_p $ we set $\Phi(Y)=\exp_p(Y)$, while
for $Y\in \Omega\setminus {\cal O}_p$ we set
$\Phi(Y)=\exp_r((Y_a-x^a)e^{a\mu}(r))  $, where $r$ is any point 
satisfying the requirements of the previous paragraph. By \eq{xeq}
$\Phi$ is well--defined (independent of the choice of $r$ in the
allowed class). $\Phi$ is smooth by smooth dependence of solutions of
ODE's with smooth coefficients upon initial values. It is injective by
eq. \eq{xeq}. The image of $\Phi$ is open in $\Omega$ because $\Phi$
is a local diffeomorphism. It is closed by Lemma \ref{L:mpx}. It
follows that $\Phi$ is surjective, hence a bijection, which
finishes the proof of the missing step of the proof in \cite{ChNad}.

\bibliographystyle{unsrt}
\bibliography{/users/piotr/prace/references/hip_bib,%
/users/piotr/prace/references/reffile,%
/users/piotr/prace/references/newbiblio,%
/users/piotr/prace/references/bibl,%
/users/piotr/prace/references/addon}

\begin{thebibliography}{10}

\bibitem{Israel:uniqueness}
W.~Israel.
\newblock Event horizons in static electrovac space-times.
\newblock {\em Commun. Math. Phys.}, 8:245--260, 1968.

\bibitem{CarterlesHouches}
B.~Carter.
\newblock Black hole equilibrium states.
\newblock In C.\ de~Witt and B.\ de~Witt, editors, {\em Black Holes}. Gordon \&
  Breach, New York, London, Paris, 1973.
\newblock Proceedings of the Les Houches Summer School.

\bibitem{Ha1}
S.W. Hawking.
\newblock Black holes in general relativity.
\newblock {\em Commun. Math. Phys.}, 25:152--166, 1972.

\bibitem{HE}
S.W. Hawking and G.F.R. Ellis.
\newblock {\em The large scale structure of space-time}.
\newblock Cambridge University Press, Cambridge, 1973.

\bibitem{Mazur}
P.~Mazur.
\newblock Proof of uniqueness of the {K}err--{N}ewman black hole solution.
\newblock {\em Jour.\ Phys.\ A: Math.\ Gen.}, 15:3173--3180, 1982.

\bibitem{Ruback}
P.\ Ruback.
\newblock A new uniqueness theorem for charged black holes.
\newblock {\em Class.\ Quantum Grav}, 5:L155--L159, 1988.

\bibitem{bunting:masood}
G.~Bunting and A.K.M.~Masood ul~Alam.
\newblock Nonexistence of multiple black holes in asymptotically euclidean
  static vacuum space-time.
\newblock {\em Gen.\ Rel.\ Grav.}, 19:147--154, 1987.

\bibitem{Sudarsky:wald}
D.\ Sudarsky and R.M.\ Wald.
\newblock Extrema of mass, stationarity and staticity, and solutions to the
  {E}instein--{Y}ang--{M}ills equations.
\newblock {\em Phys.\ Rev.}, D46:1453--1474, 1993.

\bibitem{Heusler:book}
M.~Heusler.
\newblock {\em Black hole uniqueness theorems}.
\newblock Cambridge University Press, Cambridge, 1996.

\bibitem{Chnohair}
P.T. Chru\'sciel.
\newblock ``{N}o {H}air'' {T}heorems -- folklore, conjectures, results.
\newblock In J.~Beem and K.L. Duggal, editors, {\em Differential Geometry and
  Mathematical Physics}, volume 170, pages 23--49. American Mathematical
  Society, Providence, 1994.
\newblock gr-qc/9402032.

\bibitem{Bizon:bhreview}
P.\ Bizo\'n.
\newblock Gravitating solitons and hairy black holes.
\newblock {\em Acta Physica Polonica}, B24:877--898, 1994.
\newblock gr-qc/9402016.

\bibitem{Garyonbh}
G.T. Horowitz.
\newblock The origin of black hole entropy in string theory.
\newblock UCSB preprint UCSBTH-96-07, gr-qc/9604051, 1996.

\bibitem{SudarskyNunez}
D.~N{\'u\~n}ez, H.~Quevedo, and D.~Sudarsky.
\newblock Black holes have no short hair.
\newblock {\em Physical Review Letters}, page in press, 1996.
\newblock gr-qc/9601020.

\bibitem{ChBeig2}
R.~Beig and P.T. Chru\'sciel.
\newblock The isometry groups of asymptotically flat, asymptotically empty
  space--times with timelike {ADM} four--momentum.
\newblock 1996.
\newblock preprint.

\bibitem{Ch:rigidity}
P.T. Chru\'sciel.
\newblock On rigidity of analytic black holes.
\newblock Tours preprint, 1996.

\bibitem{ChWald}
P.T. Chru\'sciel and R.M. Wald.
\newblock On the topology of stationary black holes.
\newblock {\em Class. Quantum Grav.}, 11:L147--152, 1994.

\bibitem{Waldbook}
R.M. Wald.
\newblock {\em General Relativity}.
\newblock University of Chicago Press, Chicago, 1984.

\bibitem{galloway-topology}
G.J. Galloway.
\newblock On the topology of the domain of outer communication.
\newblock {\em Class. Quantum Grav.}, 12:L99--L101, 1995.

\bibitem{Galloway:fitopology}
G.J. Galloway.
\newblock A ``finite infinity'' version of the {FSW} topological censorship.
\newblock {\em Class. Quantum Grav.}, 13:1471--1478, 1996.

\bibitem{FriedmanSchleichWitt}
J.L. Friedman, K.~Schleich, and D.M. Witt.
\newblock Topological censorship.
\newblock {\em Phys. Rev. Lett.}, 71:1486--1489, 1993.

\bibitem{Jacobson:venkatarami}
T.\ Jacobson and S.\ Venkatarami.
\newblock Topology of event horizons and topological censorship.
\newblock {\em Class. Quantum Grav.}, 12:1055--1061, 1995.

\bibitem{Woolgar:BCP}
E.~Woolgar.
\newblock Fastest curves and toroidal black holes.
\newblock In P.T. Chru\'sciel, editor, {\em Proceedings of the Minisemester on
  Mathematical Aspects of Theories of Gravitation}, Warsaw, February--March
  1996. Banach Center.

\bibitem{galloway:woolgar}
G.J. Galloway and E.~Woolgar.
\newblock The cosmic censor forbids naked topology.
\newblock preprint, gr-qc/9609007, 1996.

\bibitem{ChWald1}
P.T. Chru\'sciel and R.M. Wald.
\newblock Maximal hypersurfaces in stationary asymptotically flat space--times.
\newblock {\em Commun. Math. Phys.}, 163:561--604, 1994.
\newblock gr--qc/9304009.

\bibitem{Damour:schmidt}
T.~Damour and B.~Schmidt.
\newblock Reliability of perturbation theory in general relativity.
\newblock {\em Jour.\ Math.\ Phys.}, 31:2441--2453, 1990.

\bibitem{Simon:elvac}
W.~Simon.
\newblock Radiative {E}instein-{M}axwell spacetimes and `no--hair' theorems.
\newblock {\em Class. Quantum Grav.}, 9:241--256, 1992.

\bibitem{Horowitz}
G.T. Horowitz.
\newblock The positive energy theorem and its extensions.
\newblock In F.~Flaherty, editor, {\em Asymptotic behavior of mass and
  spacetime geometry}, volume 202 of {\em Springer Lecture Notes in Physics}.
  Springer Verlag, New York, 1984.

\bibitem{ChBeig1}
R.~Beig and P.T. Chru\'sciel.
\newblock Killing vectors in asymptotically flat space--times: I.
  {A}symptotically translational {K}illing vectors and the rigid positive
  energy theorem.
\newblock {\em Jour.\ Math.\ Phys.}, 37:1939--1961, 1996.
\newblock gr-qc/9510015.

\bibitem{Herzlich}
M.~Herzlich.
\newblock The positive mass theorem for black holes revisited.
\newblock {\em preprint}, 1996.

\bibitem{BicakSchmidt}
J.~Bi\v{c}\'{a}k and B.~Schmidt.
\newblock Asymptotically flat radiative space--times with boost--rotation
  symmetry: The general structure.
\newblock {\em Phys. Rev.}, D40:1827--1853, 1989.

\bibitem{Chandra:ellipso}
S.~Chandrasekhar.
\newblock {\em Ellipsoidal figures of equilibrium}.
\newblock Dover Publ., New York, 1969.

\bibitem{CarterCargese}
B.~Carter.
\newblock Mathematical foundation of the theory of relativistic stellar and
  black hole configurations.
\newblock In B.\ Carter and J.B.\ Hartle, editors, {\em Gravitation and
  Astrophysics}. Plenum Press, New York, 1987.

\bibitem{Weinstein3}
G.~Weinstein.
\newblock {$N$}-black hole stationary and axially symmetric solutions of the
  {E}instein/{M}axwell equations.
\newblock {\em Comm. Part. Diff. Eqs.}, 1996.
\newblock in press.

\bibitem{Chorbits}
P.T. Chru\'sciel.
\newblock On completeness of orbits of {K}illing vector fields.
\newblock {\em Class. Quantum Grav.}, 10:2091--2101, 1993.
\newblock gr-qc/9304029.

\bibitem{Nomizu}
K.~Nomizu.
\newblock On local and global existence of {K}illing vector fields.
\newblock {\em Ann. Math.}, 72:105--120, 1960.

\bibitem{VinceJimcompactCauchyCMP}
V.~Moncrief and J.~Isenberg.
\newblock Symmetries of cosmological {C}auchy horizons.
\newblock {\em Commun. Math. Phys.}, 89:387--413, 1983.

\bibitem{Bicak:podolsky}
J.~Bi\v{c}\'ak and J.~Podolsk\'y.
\newblock The global structure of {Robinson-Trautman} radiative space-times
  with cosmological constant.
\newblock 1996.
\newblock Potsdam preprint AEI-013.

\bibitem{RaczWald2}
I.~R{\'a}cz and R.~Wald.
\newblock Global extensions of spacetimes describing asymptotic final states of
  black holes.
\newblock {\em Class. Quantum Grav.}, 13:539--552, 1996.
\newblock gr-qc/9507055.

\bibitem{Gibbons:inBarrow}
G.~Gibbons.
\newblock Self-gravitating magnetic monopoles, global monopoles and black
  holes.
\newblock In J.D. Barrow, A.B. Henriques, M.T.V.T. Lago, and M.S. Longair,
  editors, {\em The Physical universe: the interface between cosmology,
  astrophysics and particle physics}, pages 110--133. Springer Verlag, Berlin,
  1991.
\newblock Springer Lecture Notes in Physics, vol. 383.

\bibitem{Majumdar}
S.D. Majumdar.
\newblock A class of exact solutions of {E}instein's field equations.
\newblock {\em Phys. Rev.}, 72:390--398, 1947.

\bibitem{Papapetrou:MP}
A.~Papapetrou.
\newblock A static solution of the equations of the gravitational field for an
  arbitrary charge distribution.
\newblock {\em Proc. Roy. Irish Acad.}, A51:191--204, 1947.

\bibitem{HartleHawking}
J.B. Hartle and S.W. Hawking.
\newblock Solutions of the {E}instein--{M}axwell equations with many black
  holes.
\newblock {\em Commun.\ Math.\ Phys.}, 26:87--101, 1972.

\bibitem{heuslerMP}
M.~Heusler.
\newblock On the uniqueness of the {P}apapetrou--{M}ajumdar metric.
\newblock 1996.
\newblock Zurich University preprint, gr-qc/9607001.

\bibitem{Heusler:ascona}
M.~Heusler.
\newblock No-hair theorems and black holes with hair.
\newblock 1996.
\newblock Proceedings of Journ{\'e}es Relativistes 1996, Ascona, Mai 96.

\bibitem{ChNad}
P.T. Chru\'sciel and N.S. Nadirashvili.
\newblock All electrovacuum {M}ajumdar--{P}apapetrou spacetimes with
  non--singular black holes.
\newblock {\em Class. Quantum Grav.}, 12:L17--L23, 1995.
\newblock gr-qc/9412044.

\bibitem{ChmassCMP}
P.T. Chru\'sciel.
\newblock On the invariant mass conjecture in general relativity.
\newblock {\em Commun.\ Math.\ Phys.}, 120:233--248, 1988.

\bibitem{CarterCMP}
B.~Carter.
\newblock {B}unting identity and {M}azur identity for non--linear elliptic
  systems including the black hole equilibrium problem.
\newblock {\em Commun.\ Math.\ Phys.}, 99:563--591, 1985.

\bibitem{Weinstein1}
G.~Weinstein.
\newblock On rotating black--holes in equilibrium in general relativity.
\newblock {\em Commun. Pure Appl. Math.}, XLIII:903--948, 1990.

\bibitem{Weinstein2}
G.~Weinstein.
\newblock The stationary axisymmetric two--body problem in general relativity.
\newblock {\em Commun. Pure Appl. Math.}, XLV:1183--1203, 1990.

\bibitem{Weinstein:trans}
G.~Weinstein.
\newblock On the force between rotating coaxial black holes.
\newblock {\em Trans. of the Amer. Math. Soc.}, 343:899--906, 1994.

\bibitem{Fletcher}
J.~Fletcher.
\newblock {\em Non--{H}ausdorff twistor spaces and the global structure of
  space--time}.
\newblock PhD thesis, Trinity College, Oxford, 1990.

\bibitem{Carter:unpublished}
B.~Carter.
\newblock unpublished.

\bibitem{Li:Tian2}
Y.~Li and G.~Tian.
\newblock Nonexistence of axially symmetric, stationary solution of {E}instein
  vacuum equation with disconnected symmetric event horizon.
\newblock {\em Manuscripta Math.}, 73:83--89, 1991.

\bibitem{LiTian:regularity}
Y.~Li and G.~Tian.
\newblock Regularity of harmonic maps with prescribed singularities.
\newblock {\em Commun. Math. Phys.}, 149:1--30, 1992.

\bibitem{ChGalloway}
P.T. Chru\'sciel and G.J. Galloway.
\newblock ``{N}owhere'' differentiable horizons.
\newblock Univ. Tours preprint, 1996.

\bibitem{Whitt}
B.~Whitt.
\newblock Israel-{W}ilson metrics.
\newblock {\em Ann. of Phys.}, 161:241--253, 1985.

\end{thebibliography}

\end{document}